\documentclass[12pt]{article}
\begin{document}
\pagestyle{plain} \setcounter{page}{1}
\begin{center}
{\large \textbf{On Quantization of the Electrical Charge--Mass}}
\vskip 0.2 true in
{\large Dmitriy M Palatnik}\footnote{e-mail: palatnik@ripco.com}
\vskip 0.2 true in
\textit{6400 N Sheridan Rd 2605, Chicago, IL 60626}
\vskip 0.2 true in

\begin{abstract}

Suggested a non-linear, non-gauge invariant model of Maxwell
equations, based on the
Kaluza-Klein theory. The spectrum of elementary charges and masses is obtained.
{}
{}

\end{abstract}

PAC numbers: 11.10.-z; 11.10.Lm

\vskip 0.3 true in  \end{center}

\date{\today}

\vskip0.5 true in
\setlength{\baselineskip}{0.33in}
\section{Introduction}

Quantization of the electrical charge and mass still remains a
challenge for modern physics.
Several attempts were made to solve this problem, e.g. \cite{1,2}.
As an alternative to these consider the following approach.
Consider fields $(g_{ab}, \tilde{A}_a)$ in four-dimensional
spacetime, where $g_{ab}$ is
the metrics and $\tilde{A}_a$ is dimensionless electromagnetic
potential, related to the
one in standard units by the rule $\tilde{A}_a = \ell A_a$, where
$\ell$ is dimensional constant.
Take the following actions for a particle and fields (electromagnetic
and gravitational):
\begin{eqnarray}\label{1.6}
S_p & = & - m_1c_*\int \sqrt{G_{ab}dx^adx^b} - m_2c_*\int \tilde{A}_adx^a\,;\\
\label{1.7}
S_f & = & - {1\over{16\pi c_* \ell^2}}\int d\Omega
\sqrt{-g}\tilde{F}^{ab}\tilde{F}_{ab}\,;\\
\label{1.8}
S_g & = & - {{c_*^3}\over{16\pi k}}\int d\Omega \sqrt{-g} R(g)\,.
\end{eqnarray}
Here $\tilde{F}_{ab} = \partial_a\tilde{A}_b -
\partial_b\tilde{A}_a$, $\tilde{F}^{ab} =
G^{ac}G^{bd}\tilde{F}_{cd}$, tensor $G^{ab}$ is inverse to $G_{ab} =
g_{ab} -\tilde{A}_a\tilde{A}_b$.
The mass of the particle is denoted by $m_1$;
`mass' $m_2 = \ell^{-1}c_*^{-2}e$ is related to the electrical charge, $e$.
The speed of light is denoted by $c_*$, and $k$ is the Newtonian constant
of gravitational interactions.
This theory may be regarded as a model of the Kaluza-Klein theory. If
$\tilde{A}_a$ were a gauge
potential and transformations $\tilde{A}_a = \tilde{A'}_a +
\partial_af$ were allowed, one
might assume due to (\ref{1.6}) that $G_{ab}$ is gauge-invariant. 
Then, from definition
of $G_{ab}$ one would obtain
that metrics $g_{ab}$ transforms by the rule $g_{ab} = g'_{ab} +
2\tilde{A'}_{(a}\partial_{b)}f
+ \partial_af\partial_bf$.
Thus, as it follows from (\ref{1.6})--(\ref{1.8}), this theory is
{\em not} gauge-invariant.
Define densities, $\mu_1$ and $\mu_2$, accordingly,
\begin{equation}\label{1.4}
m_i = \int \mu_i \sqrt{-g} dV\,, (i = 1,2.)
\end{equation}
Here $dV$ is element of three-dimensional volume.
    From the least action principle it follows that `generalized' Maxwell
equations are:
\begin{eqnarray}\label{1.9}
\tilde{F}^{ab}_{\; ;b} +
G^{ad}\tilde{F}_{dc}\tilde{F}^{bc}\tilde{A}_b & = & - 4\pi\ell^2
J^a\,;\\
\label{1.10}
\tilde{F}_{[ab;c]} & = & 0\,.
\end{eqnarray}
Here semicolon denotes the covariant derivative associated with
$g_{ab}$, i.e. equations
$g_{ab;c} = 0$ take place. The electrical current,
\begin{equation}\label{1.11}
J^a = \left( \mu_2c_*^2 -
\mu_1c_*^2\tilde{A}_bU^b
\right){{U^a}\over{U^0}}\,.
\end{equation}
Here
\begin{equation}\label{1.11.1}
  U^a = {{dx^a}\over{\sqrt{G_{cd}dx^cdx^d}}}\;.
\end{equation}
The stress-energy, entering Einstein equations for $g_{ab}$ consists
of two parts; $T_{ab}$,
pertaining to the particle, and $t_{ab}$, pertaining to electromagnetic field,
\begin{eqnarray}\label{1.12}
T^{ab} & = & \mu_1c_*^2{{U^aU^b}\over{U^0}}\,;\\
\label{1.13}
t^{ab} & = & {1\over{4\pi\ell^2}}\left( -
G^{ad}\tilde{F}_{dc}\tilde{F}^{bc} + {1\over 4}g^{ab}
\tilde{F}_{cd}\tilde{F}^{cd}\right)\,.
\end{eqnarray}
Below the attention is focused on electromagnetic interactions in
flat spacetime
($g_{ab} = \eta_{ab} =$ diag$(1, -1, -1, -1)$).

\section{Non-Relativistic Limit}

In non-relativistic limit one neglects terms of the order ${v\over
{C_*}}$, where $v$ denotes
speeds of particles. Take $x^a = (c_*t, x^{\alpha}), \alpha = 1,2,3.$ The
potential is described
by $\tilde{A}_a(x^b) = (\sin\Psi(x^b), 0, 0, 0)$, generated by the
only non-vanishing component
of the current,
\begin{equation}\label{2.1}
J^0 = \mu_2c_*^2 - \mu_1c_*^2{{\tilde{A}_0}\over{\sqrt{G_{00}}}}\,.
\end{equation}
Then (\ref{1.9}) is equivalent to
\begin{equation}\label{2.2}
\Delta\Psi = - 4\pi\ell^2 c_*^2(\mu_2\cos\Psi - \mu_1\sin\Psi)\,;
\end{equation}
here $\Delta$ denotes the Laplacian.

One may write down a class of exterior solutions, corresponding to
the charge distribution $J^0$,
\begin{equation}\label{2.3}
\tilde{A}_0(x^a) = \sin[\ell\phi(x^a) + \psi]\,.
\end{equation}
Here $\phi(x^a)$ is the Coulomb potential generated by current component
$j^0(x^a) = \cos[\ell\phi(x^a) + \psi]J^0(x^a)$, and $\psi$ is the
constant of integration.
The exterior solution for a pointlike charge, $e$, in spherical coordinates,
$x^a = (c_*t, r, \theta, \varphi)$,
\begin{equation}\label{2.4}
\tilde{A}_0 = \sin\left({{\ell e}\over r} + \psi\right)\,.
\end{equation}
Since field equations are non-linear, the superposition principle
doesn't work, and asymptotic
on spatial infinity should be the same for any matter distribution.

To find respective `gravitational' potential $G_{ab}$ and
electromagnetic potential, one has to
rescale the speed of light. Really, on spatial infinity
$ds_1^2 \equiv G_{ab}dx^adx^b = \cos^2\psi c_*^2dt^2 - dr^2 - r^2(d\theta^2
+ \sin^2\theta d\varphi^2)$,
hence, one may introduce observable speed of light, $|c|$, where $c =
c_*\cos\psi$; then metrics $H_{ab}$
is defined accordingly,
$ds_1^2 = H_{ab}dx^adx^b$, where $x^a = (ct, x^{\alpha})$.
In the case considered above,
\begin{equation}\label{2.8}
ds_1^2 = {{\cos^2\left({{\ell e}\over r} +
\psi\right)}\over{\cos^2\psi}}c^2dt^2 -r^2(d\theta^2
+ \sin^2\theta d\varphi^2)\,.
\end{equation}
For `interval' $ds_2 \equiv - \ell A_a dx^a$ one obtains,
\begin{equation}\label{2.9}
ds_2 = - (\cos\psi)^{-1}\tilde{A}_0cdt - \tilde{A}_{\alpha}dx^{\alpha}\,,
\end{equation}
where $A_a$ is electromagnetic potential in the standard units. In 
considered above
spherically-symmetric case,
\begin{equation}\label{2.10}
A_0(r) = {{\sin\left({{\ell e}\over r} + \psi\right)}\over{\ell \cos\psi}}\,.
\end{equation}
For $\ell e \ll r$ one obtains from (\ref{2.8}) and (\ref{2.10}),
\begin{eqnarray}\label{2.12}
H_{00} & = & 1 - {{2\ell e \tan\psi}\over r} + O(r^{-2})\,;\\
\label{2.13}
A_0 & = & \ell^{-1}\tan\psi + {e\over r} +  O(r^{-2})\,.
\end{eqnarray}

\section{Quantizing the Electrical Charge}

Consider flat spacetime with motionless electrical charge, described
in spherical coordinate
system. Take $\tilde{A}_a = \left[\sin{{\Phi(r)}\over r}, 0, 0,
0\right]$. Equation (\ref{1.9})
is equivalent to the following,
\begin{equation}\label{3.1}
\Phi'' = 4\pi\ell^2c_*^2 r\left( -\mu_2 \cos{{\Phi}\over r} + \mu_1
\sin {{\Phi}\over r}\right)\;;
\end{equation}
here prime denotes differentiation over $r$.

One may take the following model of the current (\ref{1.11})  for the particle.
(a) For the density take
\begin{equation}\label{3.2}
- \mu_1 + i\mu_2 = {{p^2}\over{4\pi\ell^2 c_*^2}}\exp(i\psi)
{{\arctan\sqrt{\tilde{A}_a\tilde{A}^a} + \psi}\over{
\sin\left(\arctan\sqrt{\tilde{A}_a\tilde{A}^a} + \psi\right)}}\,.
\end{equation}
Here $\tilde{A}^a = G^{ab}\tilde{A}_b$, and $p$, $\psi$ are real
normalizing constants, found
from (\ref{1.4}).
(b) For four velocity $U^a$ take
\begin{equation}\label{3.2.2}
U^a = {{\tilde{A}^a}\over{\sqrt{\tilde{A}_b\tilde{A}^b}}}\;.
\end{equation}
(c) Define boundary, $S$, of the particle by equation,
\begin{equation}\label{3.2.1}
\left(\arctan\sqrt{\tilde{A}_a\tilde{A}^a} + \psi\right)_S = 0\;.
\end{equation}
It is assumed that in the exterior of the boundary $\mu_1 + i\mu_2 = 0$.

Solving (\ref{3.1}) with (\ref{3.2}), one obtains, $\tilde{A}_- =
\sin\left(C{{\sin(pr)}\over{pr}}
- \psi\right)$ and $\tilde{A}_+ = \sin\left({a\over r} + b\right)$,
respectively interior and
exterior solutions. Here $C$, $a$, and $b$ are constants of integration.

Define boundary of the particle, as a surface of the sphere $r = R$.
Then, from (\ref{3.2.1}) it follows, (a) $\sin(pR) = 0$.
Demanding continuity of the potential together with its first
derivative, one obtains two
relations: (b) $b = - {a\over R} - \psi$; (c) $a = - CR\cos(pR)$.
    From (a) it follows that
two cases take place: (i) $a = - C{{2\pi n}\over p}$ with $R = {{2\pi
n}\over p}$ and
(ii) $a = C {{\pi(2n + 1)}\over p}$ with $R = {{\pi(2n + 1)}\over
p}$. In both cases $n = 0, 1, 2,$
     ... For the exterior potential one obtains two branches, 
corresponding to cases
(i) and (ii) above.
Assuming that both branches have the same asymptotic on spatial
infinity, one obtains,
$ C = \pi N$, $N = 0, \pm 1, \pm 2, ...$ On the other hand, demanding
that mass-charge be finite,
one has to put $N = \pm 1$. Thus, in the standard units, introducing
$e$, charge of proton, one
obtains,
\begin{eqnarray}\label{3.5}
A_{0(1)} & = & \ell^{-1}\cos^{-1}\psi \sin\left(N{{(2n)\ell
e}\over{3r}} + \psi\right)\;;\\
\label{3.6}
A_{0(2)} & = & \ell^{-1}\cos^{-1}\psi \sin\left(- N{{(2n + 1)\ell
e}\over{3r}} + \psi\right)\;.
\end{eqnarray}
Both branches originate from the following interior ($r \le R$) solution:
\begin{equation}\label{3.7}
A_{0(-)} = \ell^{-1}\cos^{-1}\psi \sin\left(\pi N
{{\sin(pr)}\over{pr}} - \psi\right)\;.
\end{equation}
Here $p = 3\pi^2\ell^{-1}e^{-1}$. The respective mass-charge density is
\begin{equation}\label{3.8}
\mu_1 - i\mu_2 = -
{{p^2\exp(i\psi)}\over{4\ell^2c_*^2}}{{v(pr)}\over{\sin(\pi
v(pr))}}\;,
\end{equation}
and it doesn't depend on $N$. Thus, one may take $N = 1$.
In (\ref{3.8}) $v(x) = {{\sin x}\over x}$.

Expanding solutions (\ref{3.5}) and (\ref{3.6}), one obtains,
\begin{eqnarray}\label{3.9}
A_{0(1)} = A_* + {e\over 3}\left({{2n}\over r}\right) -
{s\over 2}\left({{2n}\over r}\right)^2 - {w\over 6}
\left({{2n}\over r}\right)^3 + \cdots\\
\label{3.10}
A_{0(2)} = A_* - {e\over 3}\left({{2n+1}\over r}\right) -
{s\over 2}\left({{2n+1}\over r}\right)^2 + {w\over 6}
\left({{2n+1}\over r}\right)^3 + \cdots
\end{eqnarray}
Here $A_* = \ell^{-1}\tan\psi$, and the following definitions are
used for charges,
\begin{eqnarray}\label{3.11}
e & = & 3\pi^2\ell^{-1}p^{-1}\;;\\
\label{3.12}
s & = & \pi^4\ell^{-1}p^{-2}\tan\psi\;;\\
\label{3.13}
w & = & \pi^6\ell^{-1}p^{-3}\;;
\end{eqnarray}
or, expressing $p, \psi$, and $\ell$ through $e, s$, and $w$, one obtains,
\begin{eqnarray}\label{3.14}
p^2 & = & {{\pi^4}\over 3}{e\over w}\;;\\
\label{3.15}
\tan\psi & = & {{\sqrt{3}s}\over{\sqrt{ew}}}\;;\\
\label{3.16}
\ell & = & {{3\sqrt{3}}\over e}\sqrt{{w\over e}}\;.
\end{eqnarray}
The spectrum of charges consists of two parts; the first one is
$q_{2n} = {{2n}\over 3}e$, ($n = 1, 2,$ ...), and
the second one is
$q_{2n'+1} = -{{2n'+1}\over 3}e$ ($n' = 0, 1, 2,$ ...), these
correspond to quarks
and leptons.

\section{Quantizing the Mass}

Insofar, parameter $\psi$ isn't fixed yet. To fix it, one may use
(\ref{1.4}) together
with quantization rule for charge.
    From (\ref{1.4}) and (\ref{3.8}) it follows
\begin{eqnarray}\label{3.17}
m_{1(k)} - im_{2(k)} & = & - {{\pi \exp(i\psi)}\over{p \ell^2
c_*^2}}\alpha_k \;;\\
\label{3.21}
\alpha_k & = & \int_0^{\pi k} {{x^2 v(x) dx}\over {\sin{(\pi v(x))}}}\;.
\end{eqnarray}
Here $k = 2n$ for electrical charge ${{2n}\over 3}e$, and $k = 2n +
1$ for electrical
charge $-{{2n+1}\over 3}e$.
On the other hand, $m_{2(2n)} = - {{\pi^2(2n)}\over{p\ell^2 c_*^2}}$, and
$m_{2(2n+1)} =  {{\pi^2(2n+1)}\over{p\ell^2 c_*^2}}$. It follows, then,
$\sin\psi_{2n} = - {{\pi(2n)}\over{\alpha_{2n}}}$, and $\sin\psi_{2n+1} =
{{\pi(2n+1)}\over{\alpha_{2n+1}}}$.
Hence, one obtains spectrum of masses,
\begin{eqnarray}\label{3.22}
m_{2n} & = & {e\over{3\pi\ell c_*^2}}\sqrt{\alpha_{2n}^2 -
4\pi^2 n^2}\;;\; n = 1, 2, ...\\
\label{3.23}
m_{2n'+1} & = & {e\over{3\pi\ell c_*^2}}\sqrt{\alpha_{2n'+1}^2 -
\pi^2 (2n' + 1)^2}\;;\; n' = 0, 1, 2,...
\end{eqnarray}
here   masses $m_{2n}$ correspond to charges $q_{2n}$, and
masses  $m_{2n'+1}$ correspond to charges  $q_{2n'+1}$.
For the first three alphas one obtains, $\alpha_1 \approx 6.27;
\alpha_2 \approx 30.2; \alpha_3 \approx 93.5$.
The respective masses for two quarks are
$m_{-{1\over3}} = .059m_e$ and $m_{{2\over3}} = .32m_e$,
where $m_e$ is mass of an electron.

\section{Conclusion}

In this work we obtained spectrum of electrical charges,
corresponding to leptons, ($e, \mu, \tau$),
and quarks ($u, d, s, c, t, b$). It appears, that spectrum of masses 
(\ref{3.22})
and (\ref{3.23}) doesn't
describe quarks ($s, c, t, b$). On the other hand, correct spectrum
of electrical charges
allows to believe that the analogous result may follow from the
Kaluza-Klein theory without
assumption of topological closeness of the extra-dimension.

Definition of density (\ref{3.2}) is made accordingly in order to obtain the
simpliest differential equation for potential (\ref{3.1}). A more consistent
and realistic approach would be considering constant densities,
$\mu_1$ and $\mu_2$. The price to be paid  for that is to deal with
non-linear equation for potential. We thought of doing that as unworthy,
since the theory anyway is just a model of Kaluza-Klein theory,
so that the effort of solving non-linear equation for potential should be
made in frames of Kaluza-Klein.

The masses are calculated for isolated particles. In realistic situations of
many interacting particles constant $\psi$ should be found for the
complete closed system which would shift the masses (but not the
electrical charges).

\section{Acknowledgments}

I wish to thank Boris S Tsirelson for helping me with numeric computations.

\end{document}